\documentclass[12pt,preprint]{emulateapj}

\usepackage{apjfonts}

\newcommand{\skipthis}[1]{}
\newcommand{\flux}[1]{\mbox{$S_{#1}$}}

\newcommand{\um}{\mbox{$\mu$m}}
\newcommand{\Lsun}{\mbox{L$_{\odot}$}}
\newcommand{\Msun}{\mbox{M$_{\odot}$}}
\newcommand{\kms}{\hbox{km\,s$^{-1}$}}
\newcommand{\etal}{\mbox{et al.}}
\newcommand{\ammon}{NH$_{3}$}

\slugcomment{Received 2005 January 19; accepted 2005 February 18}
\journalinfo{To appear in ApJ v626, 2005 June 10}

\begin{document}

\shortauthors{BOURKE, HYLAND, \& ROBINSON}
\shorttitle{IDENTIFICATION OF HIGH MASS STAR FORMING REGIONS}

\title
{On the Identification of High Mass Star Forming Regions using IRAS:
Contamination by Low-Mass Protostars}

\author{Tyler L. Bourke}
\affil{Harvard-Smithsonian Center for
Astrophysics, 60 Garden Street, Cambridge, MA 02138
\email{tbourke@cfa.harvard.edu}}

\author{A.R. Hyland}
\affil{Deputy Vice-Chancellor's Office, James Cook
University, Townsville, QLD 4811, Australia}

\and

\author{Garry Robinson}
\affil{
School of Physical, Environmental and Mathematical Sciences, UNSW@ADFA,\\
Australian Defence Force Academy, Canberra, ACT 2600, Australia}

\begin{abstract}

We present the results of a survey of a small sample (14) of low-mass
protostars ($L_{IR} < 10^3$ \Lsun) for 6.7 GHz methanol maser emission
performed using the ATNF Parkes radio telescope.   No new masers were
discovered.  We find that the lower luminosity limit for maser emission is
near $10^3$ \Lsun, by comparison of the sources in our sample with
previously detected methanol maser sources.  We examine the IRAS properties
of our sample and compare them with sources previously observed for
methanol maser emission, almost all of which satisfy the Wood \& Churchwell
criterion for selecting candidate UCHII regions.  We find that about half
of our sample satisfy this criterion, and in addition almost all of this
subgroup have integrated fluxes between 25 and 60 \um\ that are similar to
sources with detectable methanol maser emission.  By identifying a number
of low-mass protostars in this work and from the literature that satisfy
the Wood \& Churchwell criterion for candidate UCHII regions, we show
conclusively for the first time that the fainter flux end of their sample
is contaminated by lower-mass non-ionizing sources, confirming the
suggestion by van der Walt and Ramesh \& Sridharan.

\end{abstract}


\section{Introduction}

Through their stellar winds, UV radiation, molecular outflows, and eventual
death as supernova, high mass stars ($M > 8$\Msun, or earlier than B2)
have a profound influence on their local environment and collectively on
the Galaxy as a whole.  The past 15 years has seen renewed interest in
the study of high mass star formation (HMSF).  This can be directly traced
to three main factors: (1) the use of the IRAS catalog to identify large
numbers of candidate HMSF regions, (2) the discovery of Class II methonal
maser emission and its clear association with known regions of HMSF, and
(3) improvements in both resolution and sensitivity at infrared and longer
wavelengths.  The recent realization that most stars form in dense clusters
adds a new dimension to HMSF studies.  Embedded clusters may be the basic
unit of star formation (Lada \& Lada 2003), many dominated by ionizing OB
stars, so the study of the earliest stages of HMSF is of direct importance
to star formation as a whole.

The IRAS Point Source Catalog (PSC; Beichman et~al.\ 1985) has been used
over the past $\sim$15 years to define samples of candidate HMSF regions,
in particular ultra-compact HII (UCHII) regions.  Wood \& Churchwell (1989,
hereafter WC89) showed that known embedded O-stars (i.e., UCHII regions)
occupy a tight region in IRAS color-color space, which is distinct from
other sources that also possess rising spectral energy distributions (SEDs)
in the far-infrared (FIR) and from the general population of PSC entries.
This pioneering study has served as a starting point and guide for many
subsequent studies of large numbers of HMSF regions.  In particular,
methanol maser and radio continuum surveys of sources selected using the
WC89 criterion have identified numerous UCHII regions (Schutte et~al.\
1993; van der Walt, Gaylard \& MacLeod 1995; Walsh et~al.\ 1997).   More
recently the WC89 criterion together with IRAS flux criteria and an absence
of significant radio continuum emission (i.e., no associated UCHII) has
been used to identify and study the supposedly younger high mass
protostellar objects (HMPOs; Palla et~al.\ 1991; Molinari et~al.\ 1996;
Sridharan et~al.\ 2002).  Hundreds of HMSF regions have now been identified
and detailed studies of subsamples (Walsh et~al.\ 1998, 1999, 2001;
Phillips et~al.\ 1998, Beuther et~al.\ 2002a,b,c) and selected objects
(Hunter, Phillips \& Menten 1997; Hunter \etal\ 1998, 1999; Molinari \etal\
1998; Zhang \etal\ 1999, 2001, 2002) have been made in recent years. 

%
%
\begin{deluxetable*}{lccccccccccc}
\tablecolumns{12}
\tablewidth{0pt}
\tabletypesize{\footnotesize}
\tablecaption{Properties of the low mass protostars observed for methanol
maser emission. \label{tbl-props}}
\tablehead{
\multicolumn{2}{c}{} & 
\multicolumn{4}{c}{IRAS FLUXES (Jy)} &
\multicolumn{1}{c}{DISTANCE} &
\multicolumn{1}{c}{$L_{IR}$} &
\multicolumn{4}{c}{} \\
\cline{3-6}
\colhead{NAME} & 
\colhead{ASSOCIATION} & 
\colhead{12\um} & 
\colhead{25\um} & 
\colhead{60\um} & 
\colhead{100\um} & 
\colhead{(pc)} & 
\colhead{(\Lsun)} & 
\colhead{WC89\tablenotemark{a}} & 
\colhead{[25-12]} & 
\colhead{[60-12]} & 
\colhead{log($F_{25-60}$)} 
}
\startdata
IRAS 08076-3556 & CG~30 & 0.63 & 3.73 & 18.25 & 47.54 & 
 400\tablenotemark{b} & 9 & y & 0.77 & 1.46 & 13.8 \\
IRAS 08242-5050 & HH~46/47 & 0.82 & 6.31 & 26.31 & 58.27 & 
 400\tablenotemark{b} & 13 & y & 0.89 & 1.50 & 14.0 \\
IRAS 08261-5100 & BHR 41& 0.91 & 2.50 & 4.29 & 10.91 & 
 400\tablenotemark{b} & 4 & n & 0.44 & 0.67 & 13.4 \\
IRAS 10501-5630 & \nodata & 0.25L & 1.43 & 8.1 & 36.1 & 3100\tablenotemark{c} 
 & 305 & y & $>$0.76 & $>$1.51 & 13.4 \\
IRAS 11278-5940 & \nodata & 1.32 & 1.72 & 4.6 & 37.3 & 4400\tablenotemark{d} 
 & 680 & n & 0.11 & 0.54 & 13.3 \\
IRAS 11590-6452 & BHR 71 & 0.25L & 6.53 & 77.38 & 192.9 & 
 200\tablenotemark{d} & 8 & y & $>$1.42 & $>$2.49 & 14.3 \\
IRAS 13224-5928 & BHR 87 & 1.20 & 2.35 & 8.19 & 37.21 & 
 1000\tablenotemark{e} & 40 & n & 0.29 & 0.83 & 13.5 \\
IRAS 16293-2422 & L1689 & 0.25L & 1.82 & 25.5 & 103.2 & 
 160\tablenotemark{f} & 14 & y & $>$0.86 & $>$2.00 & 15.2 \\
IRAS 16295-4452 & Sa187 & 1.2 & 5.4 & 27.1 & 52.8 & 
 700\tablenotemark{g} & 38 & y & 0.65 & 1.35 & 14.0 \\
IRAS 16289-4449 & HH~57 & 9.3 & 31.5 & 68.0 & 69.9 & 
 700\tablenotemark{g} & 120 & n & 0.53 & 0.86 & 14.5 \\
IRAS 17193-4319 & BHR 140 & 0.25L & 0.68 & 4.74 & 22.98 & 
 700\tablenotemark{h} & 10 & (y) & $>$0.43 & $>$1.28 & 13.2 \\
R CrA & \nodata & 127.6 & 228.2 & 627.0 & 1231.4 & 130\tablenotemark{i} & 45 
 & n & 0.25 & 0.69 & 15.4 \\
VV CrA & \nodata & 31.9 & 69.1 & 131.0 & 95.2 & 130\tablenotemark{i} & 10 
 & n & 0.33 & 0.61 & 14.8 \\
TY CrA & \nodata & 31.0 & 57.0 & 496.7 & 719.1 & 130\tablenotemark{i} & 21 
 & n & 0.25 & 1.20 & 15.2 \\
\enddata
\tablecomments{An ``L" in the IRAS flux columns refers to an upper limit.  
Each $L_{IR}$ value is the far infrared luminosity calculated as
described in Casoli \etal\ 1986.
Definitions for [25-12], [60-12] and Log($F_{25-60}$) are in the text.}
\tablenotetext{a}{Entried in the WC89 column refer to whether the source
satisfies the selection criterion of Wood \& Churchwell (1989), i.e., if
[25-12] $\geq$ 0.57 and [60-12] $\geq$ 1.3.  A ``y'' in parentheses
indicates a borderline case.}
\tablenotetext{b}{Vela cometary globules -- Reipurth 1983; Zealey et~al.\ 1983;
Knude \& Nielsen 2000; Woermann, Gaylard \& Otrupcek 2001}
\tablenotetext{c}{Henning \& Launhardt 1998}
\tablenotetext{d}{Bourke et al.\ 1997}
\tablenotetext{e}{R. Sutherland et al.\ 1992, private communication}
\tablenotetext{f}{Ophiuchus molecular cloud (Wilking 1992)}
\tablenotetext{g}{Reipurth et al.\ 1997}
\tablenotetext{h}{Reddening (revised distance from Bourke \etal\ 1995b)}
\tablenotetext{i}{CrA molecular cloud (Graham 1992)}
\end{deluxetable*}

Maser transitions of methanol are now well established indicators of the
early phases of HMSF (Ellingsen 2004).  Favored pump mechanisms involve
some degree of FIR radiative pumping, coupled with amplification of
background HII region continuum radiation for the lower frequency
transitions that exhibit high brightness temperatures (Sobolev, Cragg \&
Godfrey 1997; Cragg et~al.\ 2001).  The lower limit for the FIR luminosity
for which the maser process can no longer occur is not well constrained
observationally.  This is partly due to the difficulty in measuring weak
maser lines, and in establishing unbiased distances to masing sources.  The
establishment of the precise conditions under which maser processes occur,
and when they are inactive, will assist in our understanding of the
structure of star forming regions and the physical processes pertaining to
them.  In particular, the lower limit to the luminosity required to excite
the most widely observed and strongest methanol maser transition at 6.7 GHz
has not been established.  Only one targeted survey toward known low and
intermediate mass protostars has been reported by Minier \etal\ (2003).
Minier \etal\ searched for 6.7 GHz methanol maser emission toward 57
southern protostars and 65 preprotostellar (i.e., starless) dense cores.
They did not detect any maser emission toward bona fida low mass sources.
The theory of Sobolev et al. (1997) requires gas kinetic temperatures of
$\sim$30 K and dust temperatures $>$ 150 K with enhanced methanol
abundances and moderate densities to excite these masers, conditions that
might be found in some outflows from low mass protostars (e.g., BHR 71
[Bourke et~al.\ 1997, Garay et~al.\ 1998] and L1157 [Tafalla \& Bachiller
1995; Bachiller \etal\ 1995; Benedettini et~al.\ 2002]) or within low mass
hot core regions (e.g., IRAS 16293-2422 - Kuan et~al.\ 2004, Cazaux et~al.\
2003, Ceccarelli et~al.\ 1999, 2000).  Luminosities of sources selected
using the WC89 criterion are often difficult to determine because of a lack of
velocity information or to the difficulty in breaking the near/far distance
ambiguity.  Because it is now assumed a priori that methanol masers
are only associated with HMSF regions, deciding on the near/far ambiguity
is often achieved by assuming an ionising source.

In this paper we present the results of a search for 6.7 GHz methanol maser
emission from a small sample of low mass protostars that predates Minier
\etal\ 2003.   No masers were found.  The most luminous source in our
sample with $L_{IR}\sim$680 \Lsun\ is almost as luminous as the least
luminous sources showing methanol maser emission, $\sim$1000 \Lsun.  This
leads us to examine the nature of WC89 sources with low fluxes, as some of
our sample satisfy the WC89 criterion for candidate UCHII regions based
solely on their IRAS colors.  For the first time we are able to show that
some low mass protostars contaminate the low flux end of the WC89 sample, a
result that was suggested by van der Walt (1997) and Ramesh \& Sridharan
(1997) but that has not previously been confirmed observationally.  We also
estimate the contamination level of WC89 by low mass protostars.

\section{Source list}

As this was intended only as a pilot survey to examine whether any low
luminosity young stellar objects exhibit 6.7 GHz methanol maser emission,
the source list was not selected with rigor.  However, we did aim to
observe low-luminosity sources with a range of luminosities, from a few to
a few hundred solar luminosities.  Sources were selected from lists of IRAS
sources associated with southern dark molecular clouds (Persson \& Campbell
1987,1988; Persi \etal\ 1990; Bourke, Hyland \& Robinson 1995a).  Because
of their association with dark clouds, it was assumed that these sources
are relatively nearby (mostly $<$ 1kpc, and almost always $<$ 2kpc), and so
are low- to intermediate-mass objects.  The source list is given in
Table~\ref{tbl-props}.  The distance to all sources in
Table~\ref{tbl-props} was not well established at the time of the
observations.  For sources with unknown distances we selected those with
low IRAS fluxes.  Subsequent distance estimates (as indicated in
Table~\ref{tbl-props}) do indicate that the majority satisfy our luminosity
requirement.  We attempted to confirm the large distance estimates for IRAS
10501-5630 and IRAS 11278-5940 by using the Galactic rotation curve derived
by Brand \& Blitz (1993) and the values of $V_{\rm LSR}$ determined by
Bronfman, Nyman \& May 1996 ($-$13.4 \kms\ for IRAS 10501-5630) and Brand
et~al.\ 1987 ($-$13.1 \kms\ for IRAS 10501-5630 and $-21.8$ \kms\ for IRAS
11278-5940).  However it was not possible to use this method to obtain
sensible results.  As noted in Table~\ref{tbl-props}, approximately 7 out
of 14 sources satisfy the WC89 UCHII criterion, a point we will revisit.

%
%
\begin{deluxetable}{lcc}
\tablecaption{Methanol maser fluxes for ``standard" sources\label{tbl-calib}}
\tablehead{
 & \multicolumn{2}{c}{FLUX (Jy)} \\
 \cline{2-3}
\colhead{Name} & \colhead{Caswell \etal\ (1995)} & 
 \colhead{This Work} 
}
\startdata
G323.74--0.26 & 2860 & 2728 \\
G344.42+0.05 (IRAS 16586-4442) & 15 & 10 \\
G351.42+0.64 (NGC 6334F) & 3300 & 3322 \\
\enddata
\end{deluxetable}

\section{Observations}

The observations were undertaken between 1992 December 15 and 17 using the
dual--channel cooled HEMT 6.7/12.2 GHz receiver at the Parkes 64 m radio
telescope.  At the rest frequency of 6.668 GHz the half power beamwidth of
the Parkes antenna is 3\farcm3.  For all observations the Parkes
autocorrelator was split into two sections of 1024 channels, each covering
4 MHz and one of two orthogonal linear polarizations.  The velocity range
covered for each observation was 179 \kms, with a spectral resolution of
0.35 \kms\ after Hanning smoothing of the data.

On source intergration times were 20 minutes.  Reference spectra of 20
minutes taken at regular intervals throughout the observing period were
used to form quotients with the on source observations.  The resulting
spectra for the two polarizations were then averaged, baselined and Hanning
smoothed.  Because of poor weather during the observing period, system
temperatures were typically 80 -- 90 K, and the resulting 3$\sigma$
sensitivity level of the observations were typically $\sim$0.2 Jy.

Since no maser emission or absorption was detected, no ``true" flux
calibrators were observed.  However, three sources that have previous been
reported (G323.74--0.26, G344.42+0.05, G351.42+0.64) were observed as test
observations and thus provide sufficient flux calibration.  Flux
calibration of these sources was undertaken by Caswell \etal\ (1995) in the
observing period immediately following ours, and the data presented here
have been calibrated by comparison with the fluxes given in Caswell \etal\
(1995).  Table~\ref{tbl-calib}  presents a comparison of the Caswell \etal\
(1995) fluxes with those obtained here.

\section{Results and Discussion}

As indicated in Table~\ref{tbl-results}, no maser emission was detected.
The sensitivity of our survey (0.2 Jy, 3$\sigma$) is comparable to or
better than previous single-dish surveys of IRAS selected sources (e.g.,
Schutte \etal\ 1993 -- 3 Jy; van der Walt \etal\ 1995 -- 5 Jy; Walsh \etal\
1997 -- 0.3 Jy; MacLeod \etal\ 1998 -- 0.3-0.6 Jy), which have detected a
large number of methanol masers, so this is not an explanation for our
zero-detection rate, unless methanol masers from low-mass protostars are
intrinsically weaker and so below our detection threshold.  However,
subsequent studies of methanol maser emission show that this null result is
not surprising for the sources we have observed (van der Walt \etal\ 1996;
Minier \etal\ 2003).  Most large surveys for 6.7 GHz methanol maser
emission have concentrated on IRAS sources selected using the WC89 IRAS
two-color criterion for UCHII regions, [25--12] $\geq$ 0.57 and [60--12]
$\geq$ 1.3, where [$\lambda_2 - \lambda_1$] $\equiv$
log(S$_{\lambda_2}$/S$_{\lambda_1}$) and S$_\lambda$ is the IRAS flux in Jy
at wavelength $\lambda$.  Because of possible contamination from non-UCHII
regions some of these surveys also impose flux limits on their samples
(e.g., Schutte \etal\ 1993 -- \flux{60}\ $>$ 150 Jy and \flux{100}\ $>$ 400
Jy; van der Walt \etal\ 1995 and MacLeod \etal\ 1998 -- \flux{60}\ $>$ 100
Jy and \flux{100} $>$ 100 Jy), or impose other criteria such as association
with radio continuum emission indicating the presence of an HII region
(Walsh \etal\ 1997).  Approximately half of the sources in our study
satisfy the WC89 criterion (6, maybe 7, of 14, as the lower limits on the
12 \um\ flux of IRAS 17193-4319 do not exclude it from the WC89 region),
but none of these have \flux{60}\ $>$ 100 Jy as used in the major surveys
listed above.  The methanol maser survey by van der Walt \etal\ (1995) of
sources selected using the WC89 criterion with \flux{60}\ $>$ 100 Jy had a
much lower detection rate than a similar survey by Schutte \etal\ (1993),
which used \flux{60}\ $>$ 150 Jy, suggesting that the brightest WC89
sources are more likely to be associated with methanol maser emission.  The
results presented here do not contradict this viewpoint.

%
%
\begin{deluxetable}{lccc}
\tablecolumns{4}
\tablecaption{Methanol maser survey results. \label{tbl-results}}  
\tablehead{
\colhead{NAME} & 
\colhead{R.A.} & 
\colhead{Dec.} & 
\colhead{1$\sigma$ rms} \\
\colhead{} & 
\colhead{(J2000)} & 
\colhead{(J2000)} & 
\colhead{(Jy)}
}
\startdata
IRAS 08076-3556 & 08 09 31.6 & --36 04 47 & 0.06 \\
IRAS 08242-5050 & 08 25 44.4 & --51 00 05 & 0.09 \\
IRAS 08261-5100 & 08 27 37.4 & --51 10 41 & 0.09 \\
IRAS 10501-5630 & 10 52 15.5 & --56 46 28 & 0.09 \\
IRAS 11278-5940 & 11 30 09.1 & --59 57 26 & 0.07 \\
IRAS 11590-6452 & 12 01 37.1 & --65 09 06 & 0.05 \\
IRAS 13224-5928 & 13 25 48.6 & --59 42 53 & 0.06 \\
IRAS 16289-4449 & 16 32 32.3 & --44 55 38 & 0.07 \\
IRAS 16293-2422 & 16 32 22.9 & --24 28 36 & 0.07 \\
IRAS 16295-4452 & 16 33 03.5 & --44 58 30 & 0.06 \\
IRAS 17193-4319 & 17 22 55.0 & --43 22 37 & 0.05 \\
R CrA & 19 01 51.4 & --36 56 52 & 0.05 \\
VV CrA & 19 03 05.8 & --37 12 50 & 0.06 \\
TY CrA & 19 01 38.9 & --36 54 00 & 0.05  \\
\enddata
\tablecomments{No masers were detected and
1$\sigma$ rms values are listed.  Velocity coverage was typically $\pm$90
\kms\ with a channel separation of 0.35 \kms.
Units of right ascension are hours, minutes, and seconds, 
and units of declination are degrees, arcminutes, and arcseconds.}
\end{deluxetable}

%
%
\begin{figure*} 
\includegraphics[height=17cm,angle=-90]{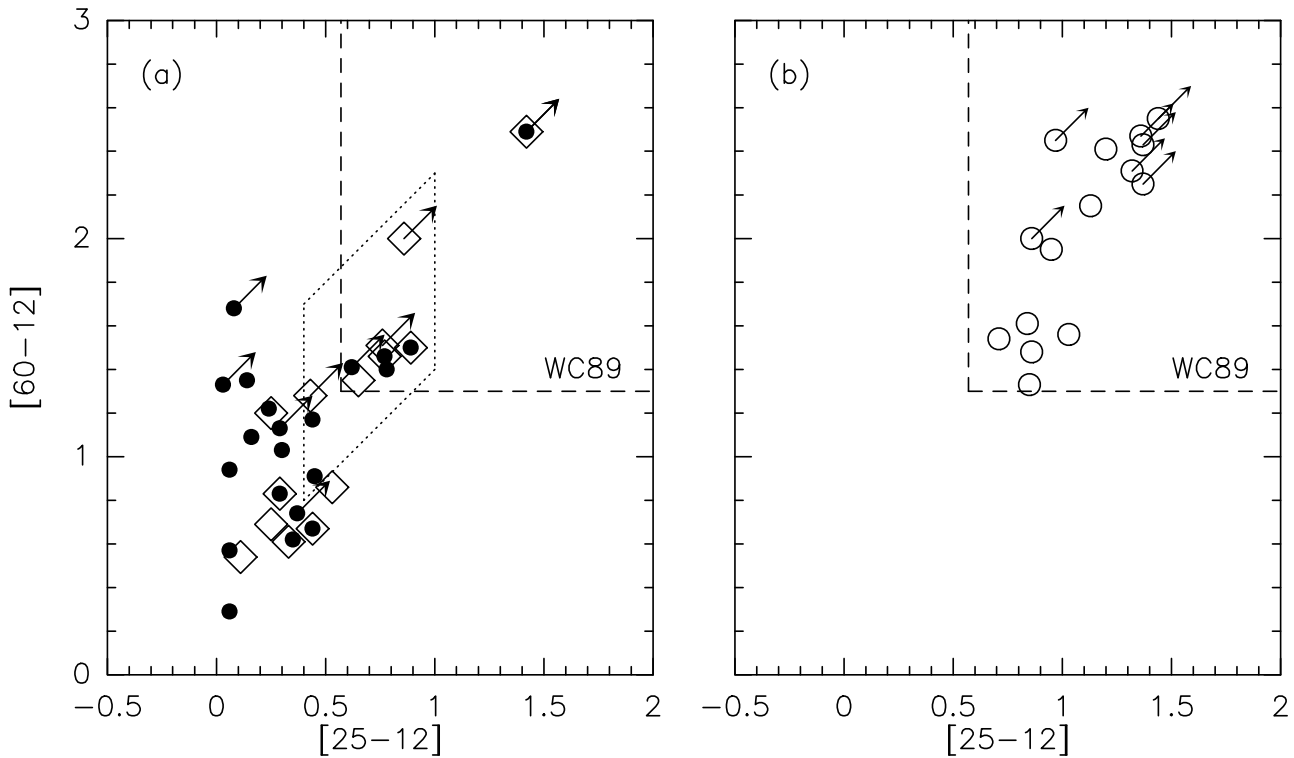}
\caption{(a) Two-color IRAS diagram showing the locations of the sources
observed for methanol maser emission in this paper ({\it open diamonds}), with
upper limits to IRAS fluxes indicated with arrows.  The filled circles
represent IRAS sources with colors of low-mass protostellar objects located
in southern Bok globules (Bourke \etal\ 1995a).   Only those sources from
Bourke \etal\ with $F_{60} > F_{25} > F_{12}$ and detected at 60 and 25
\um\ are shown.  The space occupied by UCHII regions identified by WC89 is
indicated by dashed lines, and the embedded colors region of Emerson
(1987) by the dotted parallelogram.  (b) Two-color IRAS diagram
showing the locations of known low-mass protostellar objects satisfying
WC89 and $\log F_{25-60} > 14$ (Table~\ref{tbl-lowmass}).}
\end{figure*}

\subsection{Detection toward faint IRAS sources}

The survey of van der Walt \etal\ (1996) of 241 IRAS sources satisfying
WC89 with \flux{60}\ $<$ 100 Jy for methanol maser emission at 6.7 GHz
detected only five new masers.  Because of this low detection rate van der
Walt \etal\ (1996) investigated the nature of the IRAS sources toward which
methanol masers had previously been detected.  An examination of the IRAS
flux integrated between 25 and 60 \um\ (van der Walt \etal\ 1996; Casoli
\etal\ 1986), 

\begin{eqnarray}
F_{25-60} & = & \int_{25\um}^{60\um} S(\lambda)d\lambda \nonumber \\
 & = & \left(\frac{S_{25\um}}{2} + \frac{S_{60\um}}{3.9}\right) \times
 10^{13} \,\, {\rm Jy \,\, Hz} ,
\end{eqnarray}

shows that almost all 6.7 GHz methanol masers are associated with sources
with log $F_{25-60} > 14$ Jy Hz, and that the detection rate of methanol
masers decreases rapidly with decreasing values of $F_{25-60}$.  The survey
of van der Walt \etal\ (1996) included 191 WC89 sources with 14 $<$ log
$F_{25-60}$ $<$ 14.4 that had not previously been observed, detecting the
five new masers mentioned above.  This source list included one of the
sources in our list, IRAS 11590-6452, which is clearly not an UCHII region,
but a nearby low-mass protostar (Bourke \etal\ 1997; Garay \etal\ 1998;
Bourke 2001).  In our sample seven sources have log $F_{25-60} \geq 14$ Jy Hz,
and all have log $F_{25-60} > 13$ Jy Hz, which is the lower limit for WC89
sources.

\subsection{The nature of the WC89 sample}

The probability of detecting methanol maser emission from the fainter
sources with WC89 colors is apparently quite low.  The question is why?
Are these sources more distant HMSF regions, or perhaps nearby low-mass
star forming regions that are not expected to show methanol maser emisson?
The dispersion in scale height for the WC89 sources is greatest at low
values of $F_{25-60}$, being $\sim$ 4\degr\ for log~$F_{25-60} \sim 13.5$,
and remaining constant at $\sim$ 1\degr\ for log~$F_{25-60} > 14.5$, where
most of the 6.7 GHz masers are found (van der Walt 1997).  The modeling of
the distribution of embedded ionizing stars by van der Walt (1997) led to
the conclusion that a significant fraction of the faint WC89 sources may be
low and intermediate mass (i.e., non-ionizing) nearby young stars.  This
would neatly explain the lack of 6.7 GHz maser detections for these objects
- not all WC89 objects are UCHII regions, and methanol maser emission is
  not expected from low- to intermediate-mass young stars (Sobolev \etal\
1997).  This is illustrated in Figure 1(a), which plots the [25--12] versus
[60--12] colors for the sources surveyed in this work ({\it open diamonds})
and the IRAS sources associated with Bok globules from Bourke \etal\
(1995a) that are detected at 12, 25 and 60 \um\ with
$\flux{60}>\flux{25}>\flux{12}$, and so are most likely to be low-mass
protostars.  Also shown on this figure are the regions defined by WC89 for
UCHII regions, and by Emerson (1987; see also Bourke \etal\ 1995a) for
embedded, nearby low-mass protostellar ``cores'' (Beichman \etal\ 1986).
As can be seen in the figure, a number of bona fida low luminosity sources
are found within the WC89 region, and the region of low mass cores overlaps
with the WC89 region in color-color space.  This figure clearly shows that
some low luminosity (non-ionizing) protostellar objects have IRAS colors
similar to UCHII regions and must contaminate the WC89 sample.  This point
is supported by Ramesh \& Sridharan (1997), who studied the reliability of
the WC89 criterion in the selection of UCHII regions.  Using a simple model
they estimated that $\sim$10\% of the WC89 sources have luminosities of $<$
1000 \Lsun.  

Bronfman et~al.\ (1996) surveyed 1427 sources selected using the WC89
criterion for CS $J=2-1$ emission with the Swedish-ESO Submillimetre
Telescope (SEST).  This survey was aimed at detecting dense gas (critical
density $\sim 6 \times 10^5$ cm$^{-3}$) toward these regions as a further
indicator of their protostellar nature.  They detected 843 sources (59\%)
above their sensitivity threshold.  They note that the undetected sources
are either faint IRAS sources, or seem to belong to a different population
with similar but distinct colors compared to the detected sources.  Smaller
samples of HMPO candidates selected with addition criteria (flux limits,
radio continuum) to that of WC89 (e.g., Sridharan et~al.\ 2002; Beuther
et~al.\ 2002a; Brand et~al.\ 2001) detect all of their sources in CS and
other molecular lines, regardless of distance.  This suggests that the
majority of sources not detected by Bronfman et~al.\ cannot be UCHII
regions, or even HMSF regions.

%
%
\begin{figure}[t!]
\includegraphics[height=9.9cm]{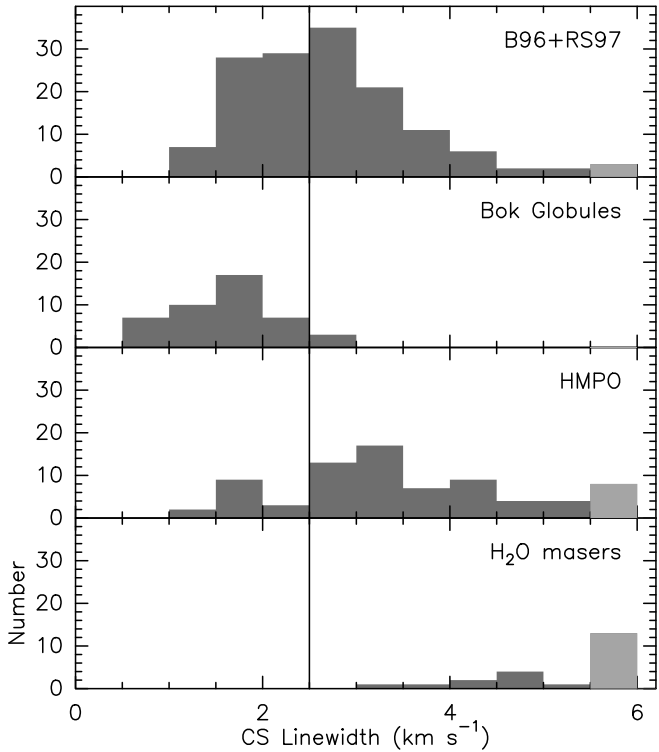}
\caption{Distribution of CS line widths for different samples of molecular
clouds as explained in the text.  The panel labeled ``B96+RS97'' shows
sources detected by Bronfman et~al.\ (1996) that are FIR faint and radio
quiet according to Ramesh \& Sridharan (1997).  The panel labeled ``Bok
Globules'' shows Bok globules detected by Launhardt et~al.\ (1998) and
Henning \& Launhardt (1998).  The panel labeled ``HMPO'' shows high mass
protostellar candidates from the surveys of Brand et~al.\ (2001) and
Beuther et~al.\ (2002a).  The panel labeled ``H$_2$O masers'' shows compact
HII regions associated with water masers detected by Juvela (1996).  The
vertical line at 2.5 \kms\ indicates the approximate division between known
low mass and high mass protostars.  In all panels, all sources with line
widths $>$5.5 \kms\ are included in the last bin (lightly shaded region),
even if their line width is $>$6 \kms.}
\end{figure}

The spectral line data allows us to investigate the nature of the detected
sources.  In particular the CS linewidth, $\Delta V$, can be used to
determine whether a given source is a candidate high mass or low mass star
forming region.  HMSF regions (star cluster forming regions without
strong ionizing sources) show larger line widths most likely because of a
greater level of turbulence.  However, there is no clear a priori
separation in linewidth between the high mass and low mass star forming
regions for any molecular tracer, and so we examined published surveys
using similar sized telescopes to estimate the typical CS linewidth of low
and high mass star forming regions.

For low mass protostars we used the Bok globule survey of Launhardt et~al.\
(1998) for northern sources, who observed 38 IRAS sources identified as low
mass protostars, and the survey of Henning \& Launhardt (1998) for southern
sources, who observed 33 IRAS sources with $L < 1000 \Lsun$.  These surveys
used the FCRAO 14 m telescope and the 15 m SEST.  They find that the
majority of the sources have $\Delta V < 2$ \kms, and essentially all have
$\Delta V < 2.5$ \kms.  For high mass protostars we have examined the CS
surveys of Brand et~al.\ (2001) and Beuther et~al.\ (2002a) for the youngest
high mass systems, and Juvela (1996) for more evolved regions (i.e.,
compact HII regions with associated water masers).  The first two surveys
generally have $\Delta V > 2$ \kms, while the Juvela sample all have
$\Delta V > 3$ \kms\ (Figure 2).  We conclude that sources in the WC89
sample with CS line widths $\Delta V \leq 2$ \kms\ are good candidates to
be regions that are not forming high mass ionising stars, or clusters
without ionizing sources.  

Larionov et~al.\ (1999) observed a mixture of high and low mass star
forming regions in CS $J=2-1$; for sources associated with Class I methanol
masers they find a mean CS linewidth $\Delta V$ = 6.2 \kms, and $\Delta V$
= 4.8 \kms\ for Class II methanol masers.  By inspection of their source
list the methanol masers are associated with previously identified HMSF
regions with radio continuum emission (ie HII regions), supporting the
result of Juvela (1996).

%
%
\begin{deluxetable*}{llccccccccc}
\tablecolumns{11}
\tablewidth{0pt}
\tabletypesize{\footnotesize}
\tablecaption{Properties of low luminosity protostars satisfying WC89 and 
log $F_{25-60} >$ 14.  \label{tbl-lowmass}}
\tablehead{
\multicolumn{2}{c}{} & 
\multicolumn{4}{c}{IRAS FLUXES (Jy)} & 
\multicolumn{1}{c}{DISTANCE} &
\multicolumn{1}{c}{$L_{IR}$} & 
\multicolumn{3}{c}{} \\
\cline{3-6}
\colhead{NAME} & 
\colhead{ASSOCIATION} & 
\colhead{12\um} & 
\colhead{25\um} & 
\colhead{60\um} & 
\colhead{100\um} & 
\colhead{(pc)} & 
\colhead{(\Lsun)} &
\colhead{[25-12]} & 
\colhead{[60-12]} & 
\colhead{log($F_{25-60}$)}
}
\startdata
03225+3034 & L1448 IRS3 & 0.6 & 5.3 & 52.2 & 374.0 & 350\tablenotemark{a} & 31 
 & 0.95 & 1.95 & 14.2 \\
03245+3002 & L1455 RNO15 & 0.2L & 4.2 & 48.8 & 82.2 & 350\tablenotemark{a} & 13 
 & $>$1.37 & 2.43 & 14.2 \\
04287+1801 & L1551 IRS5 & 9.7 & 105.0 & 356.0 & 420.0 & 140\tablenotemark{b} & 18 
 & 1.03 & 1.56 & 15.2 \\
04365+2535 & TMC 1A & 1.2 & 8.6 & 36.0 & 39.3 & 140\tablenotemark{b} & 2 
 & 0.86 & 1.48 & 14.1 \\
08194-4925 & Re4 & 0.1L & 2.3 & 29.8 & 74.6L & 400\tablenotemark{c} & $<$12 
 & $>$1.36 & $>$2.47 & 13.9 \\
08196-4931 & Re5 & 0.3 & 7.1 & 53.3 & 74.6L & 400\tablenotemark{c} & $<$19 
 & $>$1.37 & $>$2.25 & 14.2 \\
11590-6452 & BHR 71 & 0.3L & 6.5 & 77.4 & 192.9 & 200\tablenotemark{d} & 8  
 & $>$1.42 & $>$2.49 & 14.3 \\
14564-6254 & Circinus MM1 & 1.4 & 7.1 & 48.6 & 94.0 & 700\tablenotemark{e} & 63 
 & 0.71 & 1.54 & 14.2 \\
16293-2422 & L1689 & 0.25L & 1.82 & 25.5 & 103.2 & 160\tablenotemark{f} & 14 
 & $>$0.86 & $>$2.00 & 15.2 \\
18148-0440 & L483 & 0.3L & 6.9 & 89.1 & 165.5 & 200\tablenotemark{g} & 8 
 & $>$1.44 & $>$2.55 & 14.4 \\
Serpens FIRS1 & \nodata & 0.6L & 5.6 & 170.0 & 435.0 & 310\tablenotemark{h} 
 & 41 & $>$0.97 & $>$2.45 & 14.7 \\
21388+5622 & SFO 37 & 2.4 & 17.4 & 52.7 & 74.8 & 750\tablenotemark{i} & 69 
 & 0.85 & 1.33 & 14.4 \\
21391+5802 & IC 1396 & 0.6 & 8.9 & 144.6 & 425.5 & 750\tablenotemark{i} & 226 
 & 1.20 & 2.40 & 14.6 \\
22051+5848 & L1165 & 0.25L & 5.22 & 51.6 & 94 & 300\tablenotemark{j} & 11 
 & $>$1.32 & $>$2.31 & 14.2 \\
23011+6126 & Cepheus & 0.43 & 5.77 & 61.0 & 112.0 & 730\tablenotemark{k} & 76 
 & 1.13 & 2.15 & 14.3 \\
22376+7455 & L1251B & 0.8 & 5.6 & 32.3 & 66.8 & 300\tablenotemark{l} & 8 
 & 0.84 & 1.61 & 14.0 \\
\enddata
\tablecomments{Last four columns are as defined in Table 1 and
the text.  The IRAS name is given in the first column except for Serpens
FIRS1 which is confused within the IRAS beam ($\sim$1\arcmin).  An ``L" in
the IRAS flux columns refers to an upper limit.}
\tablenotetext{a}{Perseus (Herbig \& Jones 1983)}
\tablenotetext{b}{Taurus (Elias 1978)}
\tablenotetext{c}{Vela cometary globules (see Table 1)}
\tablenotetext{d}{Bourke et al.\ 1997}
\tablenotetext{e}{Bally \etal\ 1999}
\tablenotetext{f}{Ophiuchus molecular cloud (Wilking 1992)}
\tablenotetext{g}{Tafalla \etal\ 2000}
\tablenotetext{h}{Serpens (De Lara, Chavarria \& Lopez-Molina 1991)}
\tablenotetext{i}{Matthews 1979}
\tablenotetext{j}{Dobashi et~al.\ 1994}
\tablenotetext{k}{Harju et~al.\ 1993}
\tablenotetext{l}{Kun \& Prusti 1993}
\end{deluxetable*}

Plume et~al.\ (1997) observed CS $J=2-1$ and higher transitions toward
150 water masers assumed to be associated with HMSF regions (most certainly
true).  They do not list mean linewidths, but inspection of their results
indicate that 41 sources have $\Delta V <$ 3.0 \kms, and 15 (10\%) have
$\Delta V <$ 2.0 \kms.   Many of the sources with narrower linewidths are
only listed by their IRAS name, and so may be HMPOs similar to that of
Beuther et~al.\ (2002a).  Again the basic premise holds true; regions with
$\Delta V <$ 2 \kms are good candidates to be low mass star forming
regions.

In the Bronfman et~al.\ (1996) sample we find that 66 of 843 sources
detected in CS have $\Delta V < 2$ \kms.  These numbers suggest that of the
sources in WC89 that are clearly associated with molecular gas (and are
therefore likely to be star forming regions), $\sim10$\% are forming only
low mass non-ionising stars.  This suggestion is supported by the recent
survey of HMSF regions by Fa\'undez et~al.\ (2004), who
surveyed 146 of the CS detected Bronfman et~al.\ sources for 1.2 mm continuum
emission.  Of their sample $\sim9\%$ have luminosities $<10^4$ \Lsun, i.e.,
later than B0.5, and at least two are truly low mass, IRAS 08076-3556 and
IRAS 11590-6452, which are included in our list.  Ramesh \& Sridharan
(1997) divided up the WC89 sample into four classes, based on a combination
of whether they were FIR bright or faint (\flux{60} $<$ 90 Jy), and
radio-loud or radio-quiet (based on their 5~GHz emission).  For 1522
sources within $|b| < 2\fdg5$, they find $\sim27$\% are both FIR faint
and radio quiet, which are the properties expected of nearby low mass star
forming regions.  In Figure 2 we plot the distribution of CS linewidths for
those sources detected by Bronfman et al. (1996) that are also FIR faint
and radio quiet according to Ramesh \& Sridharan (1997).  This figure shows
that the majority of sources have 1.5 $< \Delta V <$ 2.9 \kms.  Of the 144
sources in common, 35 (24\%) have $\Delta V < 2$ \kms, and 64 (44\%) have
$\Delta V < 2.5$ \kms, indicating that a large fraction of faint, quiet
sources associated with dense molecular gas (confirming their star-forming
nature) have linewidths typical of low mass protostars or clusters without
ionizing sources.   If we include all sources with $\Delta V < 3.0$ \kms\
then we find 99 sources (69\%).  From these numbers we conclude that about
50\% of the FIR faint and radio quiet WC89 sources are actually low to
intermediate mass star forming regions (including non-ionizing clusters).
About one quarter of the WC89 sample are FIR faint and radio quiet
according to Ramesh \& Sridharan (1997), so we conclude for the sample as a
whole that the contamination level is $\sim10\% - 15\%$.

\subsection{Low mass, low luminosity protostars in the WB89 sample}

We have seen that attempting to select UCHII regions based solely on their
IRAS colors is insufficient to produce a clean sample.  Further criteria
are needed.  Ramesh \& Sridharan (1997) have suggested the addition of
centimeter radio continuum information, a technique used successfully and
independently by Walsh \etal\ (1997).  Surveys for methanol masers toward
UCHII candidates selected using WC89 have shown that for sources with
log~$F_{25-60} <$ 14.0 the detection rate is essentially zero, and so this
might be a useful cut-off for selecting between UCHII regions and
non-ionising sources.  Though earlier studies (van der Walt 1997; Ramesh \&
Sridharan 1997) did not actually examine the true nature of the individual
faint WC89 sources, they concluded that they are probably young embedded
non-ionizing stars.  A sample of low luminosity ($L < 1000 \Lsun$, i.e.,
non-ionizing) protostellar sources that satisfy WC89 and also have
log~$F_{25-60} >$ 14.0 are plotted in Figure 1(b), and are listed in
Table~\ref{tbl-lowmass}.  This list is not meant to be complete, but lists
a number of well known low-mass protostellar objects, including the well
studied regions in our survey associated with IRAS 08242-5050 (HH~46/47 --
Olberg \etal\ 1992; Eisl\"offel \etal\ 1994; Heathcote \etal\ 1996;
Noriega-Crespo \etal\ 2004) and IRAS 11590-6452 (BHR~71 -- Bourke \etal\
1997; Garay \etal\ 1998; Bourke 2001).  However, it should be noted that
almost all of the sources listed in Table~\ref{tbl-lowmass} have \flux{60}\
$<$ 100 Jy.  Thus, a requirement that sources satisfying WC89 also satisfy
log~$F_{25-60} >$ 14.0 and \flux{60}\ $>$ 100 Jy may select a sample with a
high percentage of UCHII regions, though it would not be a complete
Galactic census.

Two recent comprehensive studies of HMPOs have been undertaken, by
Palla and Molinari and their collaborators (Palla et~al.\ 1991; Molinari
et~al.\ 1996, 1998, 2000; hereafter the PM sample), and by Sridharan and
Beuther and their collaborators (Sridharan et~al.\ 2002; Beuther et~al.\
2002a,b,c; hereafter the SB sample).  We now briefly examine the possible
contamination level in both samples.  

In the PM sample of 260 sources, 125 satisfy WC89, which PM call ``high'',
while the rest have colors of compact molecular clouds (Richards et~al.\
1987), and are call ``low''.  In SB all 69 of their sources satisfy WC89.
Both groups attempted to discriminate against HII regions by selecting
sources without significant 5 GHz continuum emission, and against low
luminosity sources by selecting sources with \flux{60}\ $\geq$ 100 Jy (PM)
or \flux{60}\ $>$ 90 Jy and \flux{100} $>$ 500 Jy (SB).  As we have seen in
this paper, these criteria should eliminate low luminosity sources.  In
particular we have seen that the SB sample has CS linewidths significantly
larger than that of Bok globules, which only form low mass stars.  No
equivalent survey is available for PM.  They did observe a large percentage
of their list in \ammon\ (1,1) and (2,2) (163 sources), finding mean
linewidths of 1.73 \kms\ for their high sample and 1.47 \kms\ for the low
sample.  In comparison, Benson \& Myers (1989) detected \ammon\ (1,1) from
78 low mass dense cores, finding a mean linewidth of 0.43 \kms\ for those
associated with IRAS sources ($\sim50$\%), while Bourke et~al.\ (1995b)
detected 84 Bok globules in \ammon\ (1,1), with typical line widths $<$ 1
\kms.  These numbers suggest the PM sample is not likely to be contaminated
by low luminosity sources at any significant level, if at all, although
individual objects with \ammon\ linewidths $<$ 1 \kms\ should be examined
in detail.

\subsection{The lower luminosity limit for methanol maser emission}

It is difficult to determine the luminosity and hence spectral type of
protostellar sources, so it is presently unclear whether non-ionizing
sources are able to produce methanol masers.  It would be interesting to
carefully determine $L_{IR}$ for all the known methanol masers to
approximate the lower luminosity for maser emission (although IR
counterparts for a number of methanol masers are unknown), as well as the
luminosity of all those sources that been searched for methanol maser
emission with null results.  This large task is beyond the scope of this
paper.  A preliminary examination of published values of $L_{IR}$ of
methanol maser sources shows that almost all sources have $L_{IR} > 10^3$
\Lsun\ (i.e., earlier than B3), with IRAS 17463--3128 (van der Walt et al.\
1995) at 880 \Lsun\ the lowest we could find (cf. IRAS 11278--5940 with
$L_{IR} \sim 680$ \Lsun is a null detection in the current work).  If
we assume that at least 20\% of the total luminosity is not accounted for
by $L_{IR}$, then the lower luminosity limit for methanol maser emission is
$\geq 10^3$ \Lsun.

Minier \etal\ (2003) discuss their non-detection of 6.7 GHz methanol masers
from low-mass protostars in terms of the mass limit for maser emission.
However, protostellar masses are almost impossible to determine except
approximately, as they are not directly measurable.  It would be
interesting to determine the luminosity of protostars in the Minier sample
(as opposed to the starless cores), and to compare these to the values
quoted here and for faint IRAS sources showing maser emission (a task once
again beyond the scope of this paper).  Minier \etal\ claim to have
observed all southern low mass protostars known at the time of their
observations (2000), although they did not observe well studied sources
such as IRAS 08242-5050 and IRAS 11590-6452 (included in this work) which
were identified many years earlier.  This omission does not change their
conclusions; low mass protostars do not show 6.7 GHz methanol maser
emission.

\section{Summary and Conclusions}

We have undertaken a survey of a small sample of low-mass protostars for
methanol maser emission at 6.7 GHz.  No masers were detected.  This is not
surprising, as methanol maser emission is not expected to be present in the
environment of low luminosity sources (Sobolev \etal\ 1997).  The maximum
luminosity of the sources in our sample is $L_{IR} \sim$680 \Lsun, compared
to the lowest luminosity of a source showing methanol maser emission of
$L_{IR} \sim$880 \Lsun.  These results suggest that the lower
luminosity limit for methanol maser emission is $\geq 10^3$ \Lsun.

Previous studies of the IRAS properties of sources selected using the WC89
criterion and searched for 6.7 GHz methanol maser emission find that the
number of sources showing maser emission drops with decreasing IRAS flux,
and is essentially zero for sources with log~$F_{25-60} <$ 14.0.
Approximately half of our sample have log~$F_{25-60} >$ 14.0.  Most of
these also satisfy the WC89 IRAS two-color criterion for selecting
candidate UCHII regions.  These results suggest that a number of the
fainter sources satisfying WC89 with IRAS fluxes similar to known methanol
maser sources might in fact be non-ionizing lower mass protostars, as
predicted by van der Walt (1997) and Ramesh \& Sridharan (1997).  

To show this conclusively for the first time, we identified a number of
low-mass protostars in this work and from the literature that satisfy the
WC89 IRAS two-color criterion for candidate UCHII regions, and have
log~$F_{25-60} >$ 14.0, which is the observation cut off limit for 6.7 GHz
methanol maser emission.  This result clearly shows that the lower flux end
of sources satisfying this criterion is contaminated by low-mass protstars.
We estimate that the WC89 sample as a whole is contaminated by low mass
(i.e., nonionizing) protostars at least at the 10\% level, or greater.

\acknowledgments

We thank Jim Caswell for valuable advice associated with all aspects of
this program and the staff of the Parkes telescope for assistance with the
observations.  We thank an anonymous referee for insightful suggestions,
which greatly improved the final manuscript.  This project has been
supported in part by a grant from the Australian Research Council.  
T.~L.~B.\ thanks the School of Physics, ADFA, for financial assistance
during part of this study.  


\end{document}